\documentclass{article}
\usepackage{nips15submit_e,times}
\usepackage{url}
\usepackage{graphicx}
\usepackage{caption}
\usepackage{microtype}
\usepackage{enumitem}
\usepackage{amsmath,amsthm,amssymb,dsfont}
\usepackage{ifthen}
\usepackage{xspace}
\usepackage{wrapfig}
\usepackage{algorithm}
\usepackage[noend]{algorithmic}
\usepackage{color}
\definecolor{darkblue}{rgb}{0,0,0.5}
\usepackage{hyperref}
\hypersetup{
    colorlinks,
    citecolor=darkblue,
    linkcolor=blue,
    urlcolor=blue
}

\newcommand{\ind}{\mathds{1}}
\newcommand{\reals}{\mathbb{R}}

\newcommand{\bone}{\boldsymbol{1}}

\newcommand{\w}{\boldsymbol{w}}

\newcommand{\x}{\boldsymbol{x}}

\DeclareMathOperator*{\argmax}{argmax}
\DeclareMathOperator*{\argmin}{argmin}

\newcommand{\diag}{\mathrm{diag}}

\newtheorem{theorem}{Theorem}[section]
\newtheorem{lemma}{Lemma}[theorem]
\newtheorem{corollary}{Corollary}[theorem]

\newtheorem{example}{Example}[theorem]

\newcommand{\seclabel}[1]{\label{sec:#1}}
\newcommand{\secref}[1]{Section~\ref{sec:#1}}
\newcommand{\secsref}[2]{Sections~\ref{sec:#1} and \ref{sec:#2}}

\newcommand{\figlabel}[1]{\label{fig:#1}}
\newcommand{\figref}[1]{Figure~\ref{fig:#1}}

\newcommand{\tablabel}[1]{\label{tab:#1}}
\newcommand{\tabref}[1]{Table~\ref{tab:#1}}
\newcommand{\tabsref}[2]{Tables~\ref{tab:#1} and \ref{tab:#2}}

\newcommand{\eqlabel}[1]{\label{eq:#1}}
\renewcommand{\eqref}[1]{Equation~\ref{eq:#1}}

\newcommand{\lemlabel}[1]{\label{lem:#1}}
\newcommand{\lemref}[1]{Lemma~\ref{lem:#1}}

\newcommand{\thmlabel}[1]{\label{thm:#1}}
\newcommand{\thmref}[1]{Theorem~\ref{thm:#1}}

\newcommand{\alglabel}[1]{\label{alg:#1}}
\newcommand{\algref}[1]{Algorithm~\ref{alg:#1}}

\newcommand{\exlabel}[1]{\label{ex:#1}}
\newcommand{\exref}[1]{Example~\ref{ex:#1}}

\newcommand{\xor}{\triangle}
\newcommand{\shmin}{SH-min\xspace}
\newcommand{\shmax}{SH-max\xspace}
\newcommand{\submodopt}{\textsc{Submodular-Opt}\xspace}
\newcommand{\modmin}{\textsc{Modular-Min}\xspace}
\newcommand{\submodmax}{\textsc{Submodular-Max}\xspace}
\newcommand{\unionsplit}{\textsc{Union-Split}\xspace}
\newcommand{\bestB}{\textsc{Best-B}\xspace}
\newcommand{\majormin}{\textsc{Major-Min}\xspace}

\newcommand{\hamming}{HM\xspace}
\newcommand{\submod}{SM\xspace}
\newcommand{\onepart}{SP\xspace}
\newcommand{\twopart}{TP\xspace}

\providecommand{\dodraft}{false}
\newboolean{isdraft}
\setboolean{isdraft}{\dodraft}
\ifthenelse{\boolean{isdraft}}{%

{}

\newcommand{\jenny}[2]{{\color{blue}{#1 $\to$ {\bf Jenny}}: #2}}
\newcommand{\jeff}[2]{{\color{red}{#1 $\to$ {\bf Jeff}}: #2}}
\newcommand{\rahul}[2]{{\color{OliveGreen}{#1 $\to$ {\bf Rahul}}: #2}}
\newcommand{\rishabh}[2]{{\color{magenta}{#1 $\to$ {\bf Rishabh}}: #2}}
\newcommand{\bethany}[2]{{\color{green}{#1 $\to$ {\bf Bethany}}: #2}}

} {

\newcommand{\jenny}[2]{}
\newcommand{\jeff}[2]{}
\newcommand{\rahul}[2]{}
\newcommand{\rishabh}[2]{}
\newcommand{\bethany}[2]{}

}

\providecommand{\doarxiv}{true}
\usepackage{ifthen}
\newboolean{isarxiv}
\setboolean{isarxiv}{\doarxiv} 
\ifthenelse{\boolean{isarxiv}}{%
\newcommand{\arxiv}[1]{#1}
\newcommand{\notarxiv}[1]{}
}{
\newcommand{\arxiv}[1]{}
\newcommand{\notarxiv}[1]{#1}
}

\newcommand{\arxivalt}[2]{\ifthenelse{\boolean{isarxiv}}{#1}{#2}}
\newcommand{\arxivaltr}[2]{\ifthenelse{\boolean{isarxiv}}{#2}{#1}}

\title{Submodular Hamming Metrics}

\author{%
Jennifer Gillenwater$^{\dagger}$, Rishabh Iyer$^{\dagger}$, Bethany Lusch$^{\ast}$, Rahul Kidambi$^{\dagger}$, Jeff Bilmes$^{\dagger}$ \\
$^{\dagger}$ University of Washington, Dept.\ of EE, Seattle, U.S.A.\\$^{\ast}$ University of Washington, Dept. of Applied Math, Seattle, U.S.A. \\
\texttt{ \{jengi, rkiyer, herwaldt, rkidambi, bilmes\}@uw.edu}
}

\nipsfinalcopy

\begin{document}

\maketitle

\begin{abstract}
  We show that there is a largely unexplored class of functions
  (positive polymatroids) that can define proper discrete metrics over
  pairs of binary vectors and that are fairly tractable to optimize
  over.  By exploiting submodularity, we are able to give hardness
  results and approximation algorithms for optimizing over such
  metrics.  Additionally, we demonstrate empirically the effectiveness
  of these metrics and associated algorithms on both a metric
  minimization task (a form of clustering) and also a metric
  maximization task (generating diverse $k$-best lists).\looseness-1
\end{abstract}

\section{Introduction}
\seclabel{intro}
A good distance metric is often the key to an effective machine
learning algorithm.  For instance, when clustering, the distance
metric largely defines which points end up in which
clusters. Similarly, in large-margin learning, the distance between
different labelings can contribute as much to the definition of the
margin as the objective function itself.  Likewise, when constructing
diverse $k$-best lists, the measure of diversity is key to ensuring
meaningful differences between list elements.

We consider distance metrics $d: \{0,1\}^n \times \{0,1\}^n \to
\mathbb R_+$ over binary vectors, $\x \in \{0,1\}^n$.  If we define
the set $V = \{1, \ldots, n\}$, then each $\x = \bone_A$ can seen as
the characteristic vector of a set $A \subseteq V$, where $\bone_A(v)
= 1$ if $v \in A$, and $\bone_A(v) = 0$ otherwise.  For sets $A,B
\subseteq V$, with $\xor$ representing the symmetric difference, $A
\xor B \triangleq (A \setminus B) \cup (B \setminus A)$, the Hamming
distance is then\arxivalt{:\looseness-1
\begin{align}
d_H(A, B) = |A \xor B| = \sum_{i = 1}^n \bone_{A \xor B}(i) = \sum_{i = 1}^n \ind(\bone_A(i) \neq \bone_B(i)).
\end{align}}{: $d_H(A, B) = |A \xor B| = \sum_{i = 1}^n \bone_{A \xor B}(i) = \sum_{i = 1}^n \ind(\bone_A(i) \neq \bone_B(i))$. }
A Hamming distance between two vectors assumes that each entry
difference contributes value one.  Weighted Hamming distance
generalizes this slightly, allowing each entry a unique
weight. \arxivalt{Mahalanobis distance generalizes further, allowing
weighted pairwise interactions of the following form:
\begin{align}
d_M(A, B) = \bone_{A \xor B}^{\top} S \bone_{A \xor B} = \sum_{i = 1}^n \sum_{j = 1}^n S_{ij} \bone_{A \xor B}(i) \bone_{A \xor B}(j).
\end{align}
When $S$ is a positive semi-definite matrix, this type of distance is
a metric.}{The Mahalanobis distance further extends this.} For many
practical applications, however, it is desirable to have entries
interact with each other in more complex and higher-order ways than
Hamming or Mahalanobis allow.  Yet, arbitrary interactions would
result in non-metric functions whose optimization would be
intractable.  In this work, therefore, we consider an alternative
class of functions that goes beyond pairwise interactions, yet is
computationally feasible, is natural for many applications, and
preserves metricity.

Given a set function $f: 2^V \to \reals$, we can define a distortion
between two binary vectors as follows: $d_f(A,B) = f(A \xor B)$. By
asking $f$ to satisfy certain properties, we will arrive at a class of
discrete metrics that is feasible to optimize and preserves metricity.
We say that $f$ is \emph{positive} if $f(A) > 0$ whenever $A \neq
\emptyset$; $f$ is \emph{normalized} if $f(\emptyset) = 0$; $f$ is
\emph{monotone} if $f(A) \leq f(B)$ for all $A \subseteq B \subseteq
V$; $f$ is \emph{subadditive} if $f(A) + f(B) \geq f(A \cup B)$ for
all $A,B \subseteq V$; $f$ is \emph{modular} if $f(A) + f(B) = f(A
\cup B) + f(B \cap A)$ for all $A,B \subseteq V$; and $f$ is
\emph{submodular} if $f(A) + f(B) \geq f(A \cup B) + f(B \cap A)$ for
all $A,B \subseteq V$. If we assume that $f$ is positive, normalized,
monotone, and subadditive then $d_f(A,B)$ is a metric (see
\thmref{MetricPf}), but without useful computational properties.  If
$f$ is positive, normalized, monotone, and modular, then we recover
the weighted Hamming distance.  In this paper, we assume that $f$ is
positive, normalized, monotone, and \emph{sub}modular (and hence also
subadditive).  These conditions are sufficient to ensure the metricity
of $d_f$, but allow for a significant generalization over the weighted
Hamming distance.  Also, thanks to the properties of submodularity,
this class yields efficient optimization algorithms with guarantees
for practical machine learning problems.  In what follows, we will
refer to normalized monotone submodular functions as
\emph{polymatroid} functions; all of our results will be concerned
with \emph{positive polymatroids}.  We note here that despite the
restrictions described above, the polymatroid class is in fact quite
broad; it contains a number of natural choices of diversity and
coverage functions, such as set cover, facility location, saturated
coverage, and concave-over-modular functions.


Given a positive polymatroid function $f$, we refer to $d_f(A, B) =
f(A \xor B)$ as a \emph{submodular Hamming (SH) distance}. We study
two optimization problems involving these metrics (each $f_i$ is a
positive polymatroid, each $B_i \subseteq V$, and $\mathcal{C}$
denotes a combinatorial constraint):
\vspace{-1ex}
\begin{align}
\mbox{\shmin: } \min_{A \in \mathcal{C}} \sum_{i = 1}^m f_i(A \xor B_i),
\qquad \textrm{and} \qquad
\mbox{\shmax: } \max_{A \in \mathcal{C}} \sum_{i = 1}^m f_i(A \xor B_i).
\end{align}
\vspace{-2ex}

We will use $\mathcal{F}$ as shorthand for the sequence $(f_1, \ldots,
f_m)$, $\mathcal{B}$ for the sequence $(B_1, \ldots, B_m)$, and $F(A)$
for the objective function $\sum_{i = 1}^m f_i(A \xor B_i)$.  We will
also make a distinction between the \emph{homogeneous} case where all
$f_i$ are the same function, and the more general
\emph{heterogeneous} case where each $f_i$ may be distinct.  In terms
of constraints, in this paper's theory we consider only the
unconstrained ($\mathcal C = 2^V$) and the cardinality-constrained
(e.g., $|A| \geq k$, $|A| \leq k$) settings.  In general though,
$\mathcal{C}$ could express more complex concepts such as knapsack
constraints, or that solutions must be an independent set of a
matroid, or a cut (or spanning tree, path, or matching) in a graph.

Intuitively, the \shmin problem can be thought of as a
centroid-finding problem; the minimizing $A$ should be as similar to
the $B_i$'s as possible, since a penalty of $f_i(A \xor B_i)$ is paid
for each difference.  Analogously, the \shmax problem can be thought
of as a diversification problem; the maximizing $A$ should be as
distinct from all $B_i$'s as possible, as $f_i(A \xor B)$ is awarded
for each difference. Given modular $f_i$ (the weighted Hamming
distance case), these optimization problems can be solved exactly and
efficiently for many constraint types.  For the more general case of
submodular $f_i$, we establish several hardness results and offer new
approximation algorithms, as summarized in
\tabsref{theory-summary-hard}{theory-summary-alg}.  Our main contribution
is to provide (to our knowledge), the first systematic study of the
properties of submodular Hamming (SH) metrics, by showing metricity,
describing potential machine learning applications, and providing
optimization algorithms for \shmin and \shmax.

\begin{table}
\centering
\caption{Hardness for \shmin and \shmax.   UC stands for
  unconstrained, and Card stands for cardinality-constrained.  The
  entry ``open'' implies that the problem is potentially poly-time
  solvable.}
\tablabel{theory-summary-hard}
\begin{tabular}{|c|c|c|c|c|} \hline
& \multicolumn{2}{|c|}{\shmin} & \multicolumn{2}{|c|}{\shmax} \\ \cline{2-5}
& homogeneous & heterogeneous & homogeneous & heterogeneous \\ \hline
UC & Open & $4/3$ & $3/4$ & $3/4$ \\ \hline
Card & $\Omega\left(\frac{\sqrt{n}}{1 + (\sqrt{n} - 1)(1 - \kappa_f)}\right)$ & $\Omega\left(\frac{\sqrt{n}}{1 + (\sqrt{n} - 1)(1 - \kappa_f)}\right)$ & $1 - 1/e$ & $1 - 1/e$ \\ \hline 
\end{tabular}
\end{table}

\begin{table}
\centering
\caption{Approximation guarantees of algorithms for \shmin and \shmax. '-'
  implies that no guarantee holds for the corresponding pair.
  \bestB only works for the homogeneous case, while all other algorithms work in both cases.} \tablabel{theory-summary-alg}
\begin{tabular}{|c|c|c|c|c|c|} \hline
& \multicolumn{2}{|c|}{\unionsplit} & \bestB & \majormin & \textsc{Rand-Set} \\ \cline{2-6}
& UC & Card & UC & Card & UC \\ \hline
\shmin & $2$ & - & $2 - 2/m$ & $\frac{n}{1 + (n-1)(1 - \kappa_f)}$ & - \\ \hline
\shmax & $1/4$ & $1/2e$ & - & - & $1/8$ \\ \hline
\end{tabular}
\end{table}

The outline of this paper is as follows.  In \secref{applications}, we
offer further motivation by describing several applications of \shmin
and \shmax to machine learning.  In \secref{properties}, we prove that
for a positive polymatroid function $f$, the distance $d_f(A,B) = f(A
\xor B)$ is a metric.  Then, in \secsref{minimization}{maximization}
we give hardness results and approximation algorithms, and in
\secref{experiments} we demonstrate the practical advantage that
submodular metrics have over modular metrics for several real-world
applications.

\section{Applications}
\seclabel{applications}
We motivate \shmin and \shmax by showing how they occur naturally in
several applications.

\textbf{Clustering}: Many clustering algorithms, including for example
$k$-means~\cite{lloyd1982least}, use distance functions in their
optimization.  If each item $i$ to be clustered is represented by a
binary feature vector $\mathbf{b}_i \in \{0,1\}^n$, then counting the
disagreements between $\mathbf{b}_i$ and $\mathbf{b}_j$ is one natural
distance function.  Defining sets $B_i = \{v : \mathbf{b}_i(v) = 1\}$,
this count is equivalent to the Hamming distance $|B_i \xor B_j|$.
Consider a document clustering application where $V$ is the set of all
features (e.g., $n$-grams) and $B_i$ is the set of features for
document $i$.  Hamming distance has value $2$ both when $B_i \xor B_j
= \{\textrm{``submodular''}, \textrm{``synapse''}\}$ and when $B_i
\xor B_j = \{\textrm{``submodular''}, \textrm{``modular''}\}$.
Intuitively, however, a smaller distance seems warranted in the latter
case since the difference is only in one rather than two distinct
concepts.  The submodular Hamming distances we propose in this work
can easily capture this type of behavior. Given feature clusters
$\mathcal{W}$, one can define a submodular function as: \arxivalt{
\begin{equation} \eqlabel{clustering-f}
f(Y) = \sum_{W \in \mathcal{W}} \sqrt{|Y \cap W|}.
\end{equation}}{$f(Y) = \sum_{W \in \mathcal{W}} \sqrt{|Y \cap W|}.$}
Applying this with $Y = B_i \xor B_j$, if the documents' differences
are confined to one cluster, the distance is smaller than if the
differences occur across several word clusters.  In the case discussed
above, the distances are $2$ and $\sqrt{2}$.  If this submodular
Hamming distance is used for $k$-means clustering, then the
mean-finding step becomes an instance of the \shmin problem. That is,
if cluster $j$ contains documents $C_j$, then its mean takes exactly
the following \shmin form: \arxivalt{\begin{align}
    \eqlabel{clustering-objective} \mu_j \in \argmin_{A \subseteq V}
    \sum_{i \in C_j} f(A \xor B_i).
\end{align}}{$\mu_j \in \argmin_{A \subseteq V} \sum_{i \in C_j} f(A \xor B_i)$.}

\textbf{Structured prediction}: Structured support vector machines
(SVMs) typically rely on Hamming distance to compare candidate
structures to the true one.  The margin required between the correct
structure score and a candidate score is then proportional to their
Hamming distance.  Consider the problem of segmenting an image into
foreground and background.  Let $B_i$ be image $i$'s true set of
foreground pixels.  Then Hamming distance between $B_i$ and a
candidate segmentation with foreground pixels $A$ counts the number of
mis-labeled pixels.  However, both \cite{LearningMAP} and
\cite{LearningCRFs} observe poor performance with Hamming distance and
recent work by \cite{osokin2014perceptually} shows improved
performance with richer distances that are supermodular functions of
$A$.
One potential direction for further enriching image segmentation
distance functions is thus to consider non-modular functions from
within our submodular Hamming metrics class.  These functions have the
ability to correct for the over-penalization that the current distance
functions may suffer from when the same kind of difference happens
repeatedly.  For instance, if $B_i$ differs from $A$ only in the
pixels local to a particular block of the image, then current distance
functions could be seen as over-estimating the difference.  Using a
submodular Hamming function, the ``loss-augmented inference'' step in
SVM optimization becomes an \shmax problem.  More concretely, if the
segmentation model is defined by a submodular graph cut $g(A)$, then
we have: $\max_{A \subseteq V} g(A) + f(A \xor B_i)$.  (Note that
$g(A) = g(A \xor \emptyset)$.)  In fact, \cite{yu2015lovasz} observes
superior results with this type of loss-augmented inference using a
special case of a submodular Hamming metric for the task of
multi-label image classification.

\textbf{Diverse $k$-best}: For some machine learning tasks, rather
than finding a model's single highest-scoring prediction, it is
helpful to find a diverse set of high-quality predictions.  For
instance, \cite{batra2012eccv} showed that for image segmentation and
pose tracking a diverse set of $k$ solutions tended to contain a
better predictor than the top $k$ highest-scoring solutions.
Additionally, finding diverse solutions can be beneficial for
accommodating user interaction.  For example, consider the task of
selecting $10$ photos to summarize the $100$ photos that a person took
while on vacation.  If the model's best prediction (a set of $10$
images) is rejected by the user, then the system should probably
present a substantially different prediction on its second try.
Submodular functions are a natural model for several summarization
problems~\cite{linacl, tschiatschek2014learning}.  Thus, given a
submodular summarization model $g$, and a set of existing diverse
summaries $A_1, A_2, \ldots, A_{k-1}$, one could find a $k$th summary
to present to the user by solving: \arxivalt{\begin{align} A_k =
    \argmax_{A \subseteq V, |A| = \ell} g(A) + \sum_{i = 1}^{k-1} f(A
    \xor A_i).
\end{align}}{$A_k = \argmax_{A \subseteq V, |A| = \ell} g(A) + \sum_{i = 1}^{k-1} f(A \xor A_i).$}
If $f$ and $g$ are both positive polymatroids, then this constitutes
an instance of the \shmax problem.

\section{Properties of the submodular Hamming metric}
\seclabel{properties}
We next show several interesting properties of the submodular Hamming
distance.  Proofs for all theorems and lemmas can be found in the
supplementary material.  We begin by showing that any positive
polymatroid function of $A \xor B$ is a metric.  In fact, we show the
more general result that any positive normalized monotone subadditive
function of $A \xor B$ is a metric.  This result is known (see for
instance Chapter 8 of \cite{halmos1974mt}), but we provide a proof (in
the supplementary material) for completeness.

\begin{theorem} \thmlabel{MetricPf}
Let $f: 2^V \to \reals$ be a positive normalized monotone subadditive
function. Then $d_f(A, B) = f(A \xor B)$ is a metric on $A,B \subseteq
V$.
\end{theorem}
\arxiv{
\begin{proof}
Let $A, B, C \subseteq V$ be arbitrary.  We check each of the four
properties of metrics:
\begin{enumerate}
\item Proof of non-negativity: $d(A, B) = f(A \xor B) \geq 0$
  because $f$ is normalized and positive.
\item Proof of identity of indiscernibles: $d(A, B) = 0
  \Leftrightarrow f(A \xor B) = 0 \Leftrightarrow A \xor B = \emptyset
  \Leftrightarrow A = B$.  The third implication follows because of
  normalization and positivity of $f$, and the fourth follows from the
  definition of $\xor$.
\item Proof of symmetry: $d(A, B) = f(A \xor B) = f(B \xor A) = d(B,
  A)$, by definition of $\xor$.
\item Proof of the triangle inequality: First, note
  that $A \xor B \subseteq (A \xor C) \cup (C \xor B)$.  This follows
  because each element $v \in A \setminus B$ is either in $C \setminus
  B$ (true if $v \in C$) or in $A \setminus C$ (true if $v \notin C$).
  Similarly, each element $v \in B \setminus A$ is either in $C
  \setminus A$ (true if $v \in C$) or in $B \setminus C$ (true if $v
  \notin C$).  Then, because $f$ is monotone and subadditive, we have:
  \begin{equation}
    f(A \xor B) \leq f((A \xor C) \cup (C \xor B)) \leq f(A \xor C) + f(C \xor B).
  \end{equation}
\end{enumerate}
\end{proof}
} 
While these subadditive functions are metrics, their optimization is
known to be very difficult.  The simple subadditive function example
in the introduction of \cite{jegelka2011-inference-gen-graph-cuts}
shows that subadditive minimization is inapproximable, and Theorem 17
of \cite{bateni2010tr} states that no algorithm exists for subadditive
maximization that has an approximation factor better than
$\tilde{O}(\sqrt{n})$.  By contrast, submodular minimization is
poly-time in the unconstrained setting \cite{Cunningham}, and a simple
greedy algorithm from \cite{nemhauser1978mp} gives a $1 -
1/e$-approximation for maximization of positive polymatroids subject
to a cardinality constraint.  Many other approximation results are
also known for submodular function optimization subject to various
other types of constraints.  Thus, in this work we restrict ourselves to
positive polymatroids.

\begin{corollary}
Let $f: 2^V \to \reals_+$ be a positive polymatroid function.  Then
$d_f(A, B) = f(A \xor B)$ is a metric on $A,B \subseteq V$.
\end{corollary}

This restriction does not entirely resolve the question of
optimization hardness though.  Recall that the optimization in \shmin
and \shmax is with respect to $A$, but that the $f_i$ are applied to
the sets $A \xor B_i$.  Unfortunately, the function $g_B(A) = f(A \xor
B)$, for a fixed set $B$, is neither necessarily submodular nor supermodular in
$A$.  The next example demonstrates this violation of submodularity.
\begin{example}
To be submodular, the function $g_B(A) = f(A \xor B)$ must satisfy the
following condition for all sets $A_1, A_2 \subseteq V$: $g_B(A_1) + g_B(A_2) \geq g_B(A_1 \cup A_2) + g_B(A_1 \cap A_2)$.
Consider the positive polymatroid function $f(Y) = \sqrt{|Y|}$ and let
$B$ consist of two elements: $B = \{b_1, b_2\}$.  Then for $A_1 =
\{b_1\}$ and $A_2 = \{c \}$ (with $c \notin B$):
\arxivalt{\begin{align}
g_B(A_1) + g_B(A_2) = \sqrt{1} + \sqrt{3}
< 2\sqrt{2} = g_B(A_1 \cup A_2) + g_B(A_1 \cap A_2).
\end{align}}{$g_B(A_1) + g_B(A_2) = \sqrt{1} + \sqrt{3}
< 2\sqrt{2} = g_B(A_1 \cup A_2) + g_B(A_1 \cap A_2).$}
\arxiv{This violates the definition of submodularity, implying that $g_B(A)$
is not submodular.}
\exlabel{not-submod}
\end{example}
Although $g_B(A) = f(A \xor B)$ can be non-submodular, we are
interestingly still able to make use of the fact that $f$ is
submodular in $A \xor B$ to develop approximation algorithms
for \shmin and \shmax.



\section{Minimization of the submodular Hamming metric}
\seclabel{minimization}
In this section, we focus on \shmin (the centroid-finding problem).
We consider the four cases from \tabref{theory-summary-hard}: the
constrained ($A \in \mathcal{C} \subset 2^V$) and unconstrained
($A \in \mathcal{C} = 2^V$) settings, as well as the homogeneous case
(where all $f_i$ are the same function) and the heterogeneous case.
Before diving in, we note that in all cases we assume not only the
natural oracle access to the objective function $F(A) = \sum_{i = 1}^m
f_i(A \xor B_i)$ (i.e., the ability to evaluate $F(A)$ for any
$A \subseteq V$), but also knowledge of the $B_i$ (the $\mathcal{B}$
sequence).  \thmref{inapprox-woB} shows that without knowledge of
$\mathcal{B}$, \shmin is inapproximable.  In practice, requiring
knowledge of $\mathcal{B}$ is not a significant limitation; for all of
the applications described in \secref{applications}, $\mathcal{B}$ is
naturally known.

\begin{theorem}
Let $f$ be a positive polymatroid function. Suppose that the subset
$B \subseteq V$ is fixed but unknown and $g_B(A) = f(A \xor B)$.  If
we only have an oracle for $g_B$, then there is no poly-time
approximation algorithm for minimizing $g_B$, up to any polynomial
approximation factor.
\thmlabel{inapprox-woB}
\end{theorem}
\arxiv{\begin{proof}
Define $f(Y)$ as follows:
\begin{equation}
f(Y) = \begin{cases}
0 & \text{ if } Y = \emptyset \\
1 & \text{ otherwise}.
\end{cases}
\end{equation}
Then $g_B(A) = 1$ unless $A = B$.  Thus, it would take any algorithm
an exponential number of queries on $g_B$ to find $B$.
\end{proof}}

\subsection{Unconstrained setting}

Submodular minimization is poly-time in the unconstrained
setting \cite{Cunningham}.  Since a sum of submodular functions is
itself submodular, at first glance it might then seem that the sum of
$f_i$ in \shmin can be minimized in poly-time.  However, recall
from \exref{not-submod} that the $f_i$'s are not necessarily
submodular in the optimization variable, $A$.  This means that the
question of \shmin's hardness, even in the unconstrained setting, is
an open question.  \thmref{min-noC-nonhomo-hard} resolves this
question for the heterogeneous case, showing that it is NP-hard and
that no algorithm can do better than a $4/3$-approximation guarantee.
The question of hardness in the homogeneous case remains open.

\begin{theorem}
The unconstrained and heterogeneous version of \shmin is NP-hard.
Moreover, no poly-time algorithm can achieve an approximation factor
better than $4/3$.
\thmlabel{min-noC-nonhomo-hard}
\end{theorem}
\arxiv{\begin{proof}
We first show that for any graph $G = (V, E)$ it is possible to
construct $f_i$ and $B_i$ such that the corresponding sum in
the \shmin problem has minimum value if and only if $A$ is a vertex
cover for $G$.  For constants $k > 1$ and $\epsilon > 0$, let
$\gamma_1 = \frac{2^k - 1}{2^k[(2^k - 1)^{1/k} - 1]^k} + \epsilon$ and
$\gamma_2 = \frac{\gamma_1}{2^k - 1}$.  For every edge $e = (u, v) \in
E$, define two positive polymatroid functions:
\begin{equation}
f_{1e}(Y) = (\gamma_1 |Y \cap u| + \gamma_2 |Y \cap v|)^{1/k},
\quad \textrm{and} \quad
f_{2e}(Y) = (\gamma_2|Y \cap u| + \gamma_1 |Y \cap v|)^{1/k}.
\end{equation}
Let $B_{1e} = \{u\}$ and $B_{2e} = \{v\}$ and define the sum $F^c(A)$:
\begin{equation}
F^c(A) = \sum_{e \in E} h_e(A), \;\textrm{for}\;\;
h_e(A) = f_{1e}(A \xor B_{1e}) + f_{2e}(A \xor B_{2e}).
\end{equation}
The value of each term in this sum is shown in \tabref{h-vals}.  Note
that the definition of $\gamma_2$ ensures that $(\gamma_1
+ \gamma_2)^{1/k} = 2\gamma_2^{1/k}$.

\begin{table}[H]
\caption{Values of $f_{e1}$, $f_{e2}$, and their sum, $h_e$.}
\tablabel{h-vals}
\centering
\begin{tabular}{|c|c|c|c|} \hline
Case & $f_{1e}(A \xor B_{1e})$ & $f_{2e}(A \xor B_{2e})$ & $h_e(A)$ \\ \hline
$u \in A, v \notin A$    & $0$ & $(\gamma_1 + \gamma_2)^{1/k}$ & $2\gamma_2^{1/k}$ \\ 
$u \notin A, v \notin A$ & $\gamma_1^{1/k}$ & $\gamma_1^{1/k}$ & $2\gamma_1^{1/k}$ \\
$u \in A, v \in A$       & $\gamma_2^{1/k}$ & $\gamma_2^{1/k}$ & $2\gamma_2^{1/k}$ \\
$u \notin A, v \in A$    & $(\gamma_1 + \gamma_2)^{1/k}$ & $0$ & $2\gamma_2^{1/k}$ \\ \hline
\end{tabular}
\end{table}

Using \tabref{h-vals}, we can show that the minimizers of $F^c(A)$ are
exactly the set covers of $G$:
\begin{itemize}
\item Case 1---show that every vertex cover of $G$ is a minimizer of $F^c$: 
By the definition of $\gamma_2$, we know $\gamma_1 > \gamma_2$, and so
the minimum value of $F^c$ occurs when all $h_e$ are
$2\gamma_2^{1/k}$, which is clearly achievable by setting $A = V$.
Any set $A$ that is a vertex cover contains at least one endpoint of
each edge, and hence also has value $2\gamma_2^{1/k}$ for each $h_e$.
\item Case 2---show that every minimizer of $F^c$ is a vertex cover of $G$:
Suppose that $A^*$ is a minimizer of $F^c$ but not a vertex cover of
$G$.  Then there exists some uncovered edge $e = (u, v)$ with neither
endpoint in $A^*$.  Consider adding $u$ to $A^*$ to form a set $A'$.
The corresponding difference in $h_e$ value is: $h_e(A^*) - h_e(A') =
2\gamma_1^{1/k} - 2\gamma_2^{1/k} > 0$.  The difference in $h$-value
for each other edge $e'$ that touches $u$ is similarly
$2\gamma_1^{1/k} - 2\gamma_2^{1/k}$ if $e'$ is uncovered in $A^*$, or
$0$ if $e'$ is covered by $A^*$.  All other $h$-values remain
unchanged.  Thus, $F^c(A') < F^c(A^*)$, contradicting the assumption
that $A^*$ is a minimizer of $F^c$.
\end{itemize}

Borrowing from \cite{Goel09}'s Theorem 3.1, we now define a particular
graph and two additional positive polymatroid functions.  Consider the
bipartite graph $G = (V_1 \cup V_2, E)$ where $|V_1| = |V_2| = r$ and
the edge set consists of $r$ edges that form a perfect matching of $A$
to $B$.  Let $R$ be a random minimum-cardinality vertex cover of $G$.
Define the following two functions:
\begin{equation}
f_0^a(Y) = \min\left\{|Y|, r\right\} \qquad
f_0^b(Y) = \min\left\{|Y \cap \bar{R}| + \min\left\{|Y \cap R|, \frac{(1 + \delta)r}{2}\right\}, r\right\}
\end{equation}
where $\delta$ is set so that $2/(1 + \delta) = 2
- \epsilon$.  \cite{Goel09} shows that, knowing $G$ but given only
value-oracle access to the $f_0$, no poly-time algorithm can
distinguish between $f_0^a$ and $f_0^b$.  Moreover, if restricted to
vertex cover solutions, it is easy to see that the function $f_0^a$ is
minimized on any of the $2^r$ possible vertex covers, for which it has
value $r$, while the function $f_0^b$ is minimized on the set $Y = R$,
for which it has value $\frac{(1 + \delta)r}{2}$.  The ratio of these
minimizers is $2 - \epsilon$, which allows \cite{Goel09} to show that
no poly-time algorithm can achieve a $(2 - \epsilon)$-approximation
for the minimum submodular vertex cover problem.

Now, instead of explicitly restricting to vertex cover solutions,
consider unconstrained minimization on $F^a(A) = f_0^a(A) + F^c(A)$
and $F^b(A) = f_0^b(A) + F^c(A)$.  Since $f_0^a$ and $f_0^b$ cannot be
distinguished in poly-time, neither can $F^a$ and $F^b$.  We
can also show that: (1) any minimizer of $F^a(A)$ or $F^b(A)$ must be
a vertex cover, and (2) the ratio of the corresponding vertex cover
minimizers is $4/3$.

\begin{itemize}
\item Show $F^a$'s minimizers are vertex covers: Suppose that $A^*$ is a
minimizer of $F^a$ but not a vertex cover of $G$.  Then there exists
some uncovered edge $e = (u, v)$ with neither endpoint in $A^*$.
Consider adding $u$ to $A^*$ to form a set $A'$.  As shown above, the
corresponding difference in $F^c$ value is $2\gamma_1^{1/k} -
2\gamma_2^{1/k}$.  The difference $f_0^a(A^*) - f_0^a(A')$ is $-1$ if
$|A^*| < r$ and $0$ otherwise.  Thus, all we need is for
$2\gamma_1^{1/k} - 2\gamma_2^{1/k}$ to be $> 1$.  Plugging in the
definition of $\gamma_1$ and $\gamma_2$, this inequality can be seen
to hold for all $k > 1$.  Thus, overall $F^a(A') < F^a(A^*)$,
contradicting the assumption that $A^*$ is a minimizer of $F^a$.

\item Show $F^b$'s minimizers are vertex covers: The reasoning here is
analogous to the $F^a$ case; the difference $f_0^b(A^*) - f_0^b(A')$
is always $> -1$, since adding a single node can never change the
$f_0^b$ value by more than $1$.

\item $F^a$'s minimum value: Any vertex cover $A$ includes at least $r$
nodes and thus has value $f_0^a(A) = r$.  Since there are $r$ edges
total, $F^c(A) = 2r\gamma_2^{1/k}$ for a vertex cover.  Combining
these we see that $F^a$ has minimum value $r(1 + 2\gamma_2^{1/k})$.

\item $F^b$'s minimum value: The vertex cover consisting of the set $R$
minimizes $f_0^b$: $f_0^b(R) = \frac{(1 + \delta)r}{2}$.  Thus, the
minimum $F^b$ value is $r\left(\frac{(1 + \delta)}{2} +
2\gamma_2^{1/k}\right)$.
\end{itemize}

Letting $k \to \infty$, we have that $\gamma_2^{1/k} \to 1/2$.  Thus,
in the limit the as $k \to \infty$, the ratio of minimizers is: $2 /
(\frac{(1 + \delta)}{2} + 1) = \frac{4}{3 + \delta}$.  Plugging in the
definition of $\delta$ from above, the ratio in terms of $\epsilon$
is: $\frac{4 - 2\epsilon}{3 - \epsilon} > \frac{4}{3}
- \frac{2\epsilon}{3 - \epsilon} = \frac{4}{3} - o(1)$.
\end{proof}}

Since unconstrained \shmin is NP-hard, it makes sense to consider
approximation algorithms for this problem.  We first provide a simple
$2$-approximation, \unionsplit (see \algref{union-split}).  This
algorithm splits $f(A \xor B) = f((A \setminus B) \cup (B \setminus
A))$ into $f(A \setminus B) + f(B \setminus A)$, then applies standard
submodular minimization (see e.g. \cite{Fuji05}) to the split
function.  \thmref{Fsplit} shows that this algorithm is a
$2$-approximation for \shmin.  It relies on \lemref{Fsplit}, which we
state first.

\begin{lemma} \lemlabel{Fsplit}
Let $f$ be a positive monotone subadditive function. Then, for any $A,
B \subseteq V$:
\begin{equation}
f(A \xor B) \leq f(A \setminus B) + f(B \setminus A)  \leq 2f(A \xor B).
\end{equation}
\end{lemma}
\arxiv{\begin{proof}
The upper bound follows from the definition of $\xor$ and the fact
that $f$ is subadditive:
\begin{equation}
f(A \xor B) =
f((A \setminus B) \cup (B \setminus A)) \leq
f(A \setminus B) + f(B \setminus A).
\end{equation}
The lower bound on $2f(A \xor B)$ follows due to the monotonicity of
$f$: $f(A\setminus B) \leq f(A \xor B)$ and $f(B \setminus A) \leq
f(A \xor B)$.  Summing these two inequalities gives the bound.
\end{proof}}

\begin{theorem}
\unionsplit is a $2$-approximation for unconstrained \shmin.
\thmlabel{Fsplit}
\end{theorem}
\arxiv{\begin{proof}
An \shmin instance seeks the minimizer of $F(A) = \sum_{i=1}^m
f_i(A \xor B_i)$.  Define $\bar{F}(A) = \sum_{i=1}^m \left[
f_i(A \setminus B_i) + f_i(B_i \setminus A) \right]$.
From \lemref{Fsplit}, we see that $\min_A \bar{F}(A)$ is a
$2$-approximation for $\min_A F(A)$ (any submodular function is also
subadditive).  Thus, if $\bar{F}$ can be minimized exactly, the result
is a $2$-approximation for \shmin.  Exact minimization of $\bar{F}$ is
possible because $\bar{F}$ is submodular in $A$.  The submodularity of
$\bar{F}$ follows from the fact that submodular functions are closed
under restriction, complementation, and addition
(see \cite{bach2011corr}, page 9).  These closure properties imply
that, for each $i$, $f_i(A \setminus B_i)$ and $f_i(B_i \setminus A)$
are both submodular in $A$, as is their sum.
\end{proof}}

\arxiv{Note that \unionsplit's $2$-approximation
bound is tight; there exists a problem instance where exactly a factor
of $2$ is achieved.  More concretely, consider $V = \{1, 2\}$, $B_1
= \{1\}$, $B_2 = \{2\}$, and $f_1(Y) = f_2(Y) = |Y|^{(1/\alpha)}$ for
$\alpha > 1$.  Then according to the $F'$ passed to \submodopt, all
solutions have value $2$.  Yet, under the true $F$ the solutions
$\{1\}$ and $\{2\}$ have the better (smaller) value $2^{(1/\alpha)}$.
Letting $\alpha \rightarrow \infty$, the quantity $2^{(1/\alpha)}$
approaches $1$, making the ratio between the correct solution and the
one given by \unionsplit possibly as large as $2$.}

Restricting to the homogeneous setting, we can provide a different
algorithm that has a better approximation guarantee than \unionsplit.
This algorithm simply checks the value of $F(A) = \sum_{i=1}^m
f(A \xor B_i)$ for each $B_i$ and returns the minimizing $B_i$.  We
call this algorithm \bestB (\algref{bestB}).  \thmref{bestB} gives the
approximation guarantee for \bestB.  This result is
known \cite{AlgStringsTreesSqns}, as the proof of the guarantee only
makes use of metricity and homogeneity (not submodularity), and these
properties are common to much other work.  We provide the proof in our
notation for completeness though.

\begin{theorem}
For $m = 1$, \bestB exactly solves unconstrained \shmin.  For $m >
1$, \bestB is a $\left(2 - \frac{2}{m}\right)$-approximation for
unconstrained homogeneous \shmin.
\thmlabel{bestB}
\end{theorem}
\arxiv{\begin{proof}
Define $F(A) = \sum_{i = 1}^m f_i(A \xor B_i)$, for $f_i$ positive
polymatroid.  Since each $f_i$ is normalized and positive, each is
minimized by $\emptyset$: $f_i(\emptyset) = 0$.  Thus, any given
$f_i(A \xor B_i)$ is minimized by setting $A = B_i$.  For $m = 1$,
this implies that \shmin is exactly solved by setting $A = B_1$.

Now consider $m > 1$ and the homogeneous setting where there is a
single $f$: $f_i = f\; \forall i$.  By \thmref{MetricPf}, $f(A \xor
B_i)$ is a metric, so it obeys the triangle inequality:
\begin{equation}
f(A \xor B_i) + f(A \xor B_j) \geq f(B_i \xor B_j) \quad \forall i,j.
\end{equation}
Fixing some $i$ and summing this inequality over all $j \neq i$:
\begin{equation}
\sum_{j \neq i} [f(A \xor B_i) + f(A \xor B_j)]
\geq \sum_{j \neq i} f(B_i \xor B_j) = \sum_{i = 1}^m f(B_i \xor B_j)
\end{equation}
where the last equality is due to the fact that polymatroids are
normalized: $f(B_i \xor B_i) = f(\emptyset) = 0$.  Regrouping terms,
$f(A \xor B_i)$ is independent of $j$, so it can be pulled out of the
summation:
\begin{equation}
(m - 2)f(A \xor B_i) + \sum_{j = 1}^m f(A \xor B_j)
\geq \sum_{j = 1}^m f(B_i \xor B_j).
\end{equation}
Notice that $\sum_{j = 1}^m f(A \xor B_j)$ is exactly $F(A)$ and
$\sum_{j = 1}^m f(B_i \xor B_j)$ is $F(B_i)$.  Substituting in this
notation and summing over all $i$:
\begin{equation}
\sum_{i = 1}^m [(m - 2)f(A \xor B_i) + F(A)]
\geq \sum_{i = 1}^m F(B_i).
\end{equation}
On the left-hand side we can again replace the sum with $F(A)$,
yielding: $2(m - 1)F(A) \geq \sum_{i = 1}^m F(B_i)$.  Since a sum over
$m$ items is larger than $m$ times the minimum term in the sum, the
remaining sum here can be replaced by a min:
\begin{equation}
2(m - 1)F(A) \geq m \min_{i \in \{1, \ldots, m\}} F(B_i).
\end{equation}
The left-hand size is exactly what the \bestB algorithm computes, and
hence the minimizing $B_i$ found by \bestB is a $\left(2 -
2/m\right)$-approximation for unconstrained homogeneous \shmin.
\end{proof}}

\arxiv{
Note that as a corollary of this result, in the case when $m = 2$, the
optimal solution for unconstrained homogeneous \shmin is to take the
best of $B_1$ and $B_2$.  Also note that since \unionsplit's
$2$-approximation bound is tight, \bestB is theoretically better in
terms of worst-case performance in the unconstrained setting.
However, \unionsplit's performance on practical problems is often
better than the \bestB's, as many practical problems do not hit upon
this worst case.  For example, consider the case where $V
= \{1,2,3\}$, $f$ is simply cardinality, $f(A) = |A|$, and each $B_i$
consists of two items: $B_1 = \{1,2\}, B_2 = \{1,3\}, B_3 = \{2,3\}$.
Then the best $B_i$ has $F$-value $4$, while the set $\{1,2,3\}$ found
by \unionsplit has a lower (better) $F$-value of $3$.}


\begin{figure*}
\centering
\begin{minipage}[t]{0.5\linewidth}
\begin{algorithm}[H]
\caption{\unionsplit}
\alglabel{union-split}
\begin{algorithmic}
\STATE {\bfseries Input}: $\mathcal{F}$, $\mathcal{B}$
\STATE Define $f_i'(Y) = f_i(Y \setminus B_i) + f_i(B_i \setminus Y)$
\STATE Define $F'(Y) = \sum_{i = 1}^m f_i'(Y)$
\STATE {\bfseries Output}: \submodopt($F'$)
\end{algorithmic}
\end{algorithm}
\vspace{-0.25in}
\begin{algorithm}[H]
\caption{\bestB}
\alglabel{bestB}
\begin{algorithmic}
\STATE {\bfseries Input}: $F$, $\mathcal{B}$
\STATE $A \leftarrow B_1$
\FOR{$i = 2, \ldots, m$}
  \STATE {\bfseries if} $F(B_i) < F(A)$: $A \leftarrow B_i$
\ENDFOR
\STATE {\bfseries Output}: $A$
\end{algorithmic}
\end{algorithm}
\end{minipage}
\hspace{0.2in}
\begin{minipage}[t]{0.4\linewidth}
\begin{algorithm}[H]
\caption{\majormin}
\alglabel{major-min}
\begin{algorithmic}
\STATE {\bfseries Input}: $\mathcal{F}$, $\mathcal{B}$, $\mathcal{C}$
\STATE $A \leftarrow \emptyset$
\REPEAT
  \STATE $c \leftarrow F(A)$
  \STATE Set $\w_{\hat{F}}$ as in \eqref{w-hat-F}
  \STATE $A \leftarrow$ \modmin($\w_{\hat{F}}$, $\mathcal{C}$)
\UNTIL{$F(A) = c$}
\STATE {\bfseries Output}: $A$
\end{algorithmic}
\end{algorithm}
\end{minipage}
\end{figure*}

\subsection{Constrained setting}

In the constrained setting, the \shmin problem becomes more difficult.
Essentially, all of the hardness results established in existing work
on constrained submodular minimization applies to the
constrained \shmin problem as well.  \thmref{cardconst-hard} shows
that, even for a simple cardinality constraint and identical $f_i$
(homogeneous setting), not only is \shmin NP-hard, but also it is hard
to approximate with a factor better than $\Omega(\sqrt{n})$.

\begin{theorem} \thmlabel{cardconst-hard}
Homogeneous \shmin is NP-hard under cardinality constraints.
Moreover, no algorithm can achieve an approximation factor better than
$\Omega\left(\frac{\sqrt{n}}{1 + (\sqrt{n} - 1)(1
- \kappa_f)}\right)$, where $\kappa_f = 1 - \min_{j \in V} \frac{f(j |
V \setminus j)}{f(j)}$ denotes the curvature of $f$.  This holds even
when $m = 1$.
\end{theorem}
\arxiv{\begin{proof}
Let $m = 1$ and $B_1 = \emptyset$.  Then under cardinality
constraints, \shmin becomes $\min_{A: |A| \geq k} f(A)$.  Corollary
5.1 of \cite{curvaturesubmodular} establishes that this problem is
NP-hard and has a hardness of $\Omega(\frac{\sqrt{n}}{1 + (\sqrt{n} -
1)(1 - \kappa_f)})$.
\end{proof}}

We can also show similar hardness results for several other
combinatorial constraints including matroid constraints, shortest
paths, spanning trees, cuts, etc.~\cite{curvaturesubmodular,
Goel09}.  Note that the hardness established
in \thmref{cardconst-hard} depends on a quantity $\kappa_f$, which is
also called the \emph{curvature} of a submodular
function~\cite{vondrak2010submodularity, curvaturesubmodular}.
Intuitively, this factor measures how close a submodular function is
to a modular function.  The result suggests that the closer the
function is being modular, the easier it is to optimize.  This makes
sense, since with a modular function, \shmin can be exactly minimized
under several combinatorial constraints.  To see this for the
cardinality-constrained case, first note that for modular $f_i$, the
corresponding $F$-function is also modular.  \lemref{modular-F}
formalizes this.

\begin{lemma} \lemlabel{modular-F}
If the $f_i$ in \shmin are modular, then $F(A) = \sum_{i = 1}^m
f_i(A \xor B_i)$ is also modular.
\end{lemma}
\arxiv{\begin{proof}
Any normalized modular function $f_i$ can be represented as a vector
$\w_i \in \reals^n$, such that $f_i(Y) = \sum_{j \in Y} w_i(j) =
\w_i^{\top} \bone_Y$.  With $Y = A \xor B_i$, this can be written:
\begin{align}
f_i(A \xor B_i) =& \w_i^{\top} \left[\bone_{B_i} + \diag(\bone_{V \setminus B_i} - \bone_{B_i}) \bone_A\right] \\
=& \sum_{j \in B_i} w_i(j) + \sum_{j \in A} (-1)^{\ind(j \in B_i)} w_i(j).
\end{align}
Summing over $i$ and letting $C = \sum_{i = 1}^m \sum_{j \in B_i}
 w_i(j)$ represent the part that is constant with respect to $A$, we
 have:
\begin{equation}
F(A) = C + \sum_{j \in A} (-1)^{\ind(j \in B_i)} w_i(j).
\end{equation}
Thus, $F$ can be represented by offset $C$ and vector
$\w_F \in \reals^n$ such that $F(A) = C + \sum_{j \in A} w_F(j)$, with
entries $w_F(j) = \sum_{i = 1}^m (-1)^{\ind(j \in B_i)} w_i(j)$.  This
is sufficient to prove modularity.  (For optimization purposes, note
that $C$ can be dropped without affecting the solution to \shmin.)
\end{proof}}

Given \lemref{modular-F}, from the definition of modularity we know
that there exists some constant $C$ and vector $\w_F \in \reals^n$,
such that $F(A) = C + \sum_{j \in A} w_F(j)$.  From this
representation it is clear that $F$ can be minimized subject to the
constraint $|A| \geq k$ by choosing as the set $A$ the items
corresponding to the $k$ smallest entries in $w_F$.  Thus, for modular
$f_i$, or $f_i$ with small curvature $\kappa_{f_i}$, such constrained
minimization is relatively easy.

Having established the hardness of constrained \shmin, we now turn to
considering approximation algorithms for this problem.  Unfortunately,
the \unionsplit algorithm from the previous section requires an
efficient algorithm for submodular function minimization, and no such
algorithm exists in the constrained setting; submodular minimization
is NP-hard even under simple cardinality
constraints~\cite{svitkina2008submodular}\arxiv{ (although
see \cite{nagano2011} that shows it is possible to get solutions for a
subset of the cardinality constraints)}.\ Similarly, the \bestB
algorithm breaks down in the constrained setting; its guarantees carry
over only if all the $B_i$ are within the constraint set
$\mathcal{C}$.  Thus, for the constrained \shmin problem we instead
propose a majorization-minimization
algorithm.  \thmref{major-min-bound} shows that this algorithm has an
$O(n)$ approximation guarantee, and \algref{major-min} formally
defines the algorithm.

Essentially, \majormin proceeds by iterating the following two steps:
constructing $\hat{F}$, a modular upper bound for $F$ at the current
solution $A$, then minimizing $\hat{F}$ to get a new $A$.  $\hat{F}$
consists of
superdifferentials \cite{jegelka2011-nonsubmod-vision,rkiyersubmodBregman2012}
of $F$'s component submodular functions.  We use the
superdifferentials defined as ``grow'' and ``shrink''
in \cite{rkiyersemiframework2013}.  Defining sets $S,T$ as $S =
V \setminus j, T = A \xor B_i$ for ``grow'', and $S = (A \xor
B_i) \setminus j, T = \emptyset$ for ``shrink'', the $\w_{\hat{F}}$
vector that represents the modular $\hat{F}$ can be written:
\begin{equation}
\eqlabel{w-hat-F}
w_{\hat{F}}(j) = \sum_{i = 1}^m
\begin{cases} f_i(j \mid S) \;\textrm{if}\; j \in A \xor B_i \\
 f_i(j \mid T) \;\textrm{otherwise},
\end{cases}
\end{equation}
where $f(Y \mid X) = f(Y \cup X) - f(X)$ is the gain in $f$-value when
adding $Y$ to $X$.  We now state the main theorem characterizing
algorithm \majormin's performance on \shmin.

\begin{theorem} \thmlabel{major-min-bound}
\majormin is guaranteed to improve the objective value, $F(A) = \sum_{i = 1}^m
f_i(A \xor B_i)$, at every iteration.  Moreover, for any constraint
over which a modular function can be exactly optimized, it has a
$\left(\max_i \frac{|A^* \xor B_i|}{1+(|A^* \xor
B_i|-1)(1-\kappa_{f_i}(A^* \xor B_i))}\right)$ approximation
guarantee, where $A^*$ is the optimal solution of \shmin.
\end{theorem}
\arxiv{\begin{proof}
We first define the full ``grow'' and ``shrink'' superdifferentials:
\begin{align}
\eqlabel{grow-superdiff}
m^f_{A, 1}(Y) &\triangleq f(A) - \!\!\
 \sum_{j \in A \setminus Y } f(j \mid V \setminus j) + \!\!\!\!
\sum_{j \in Y \setminus A} f(j \mid A), \quad \textrm{and} \\
\eqlabel{shrink-superdiff}
m^f_{A, 2}(Y) &\triangleq f(A) - \!\!\!\!
\sum_{j \in A \setminus Y} f(j \mid A \setminus j) + \!\!\!\!
\sum_{j \in Y \setminus A} f(j \mid \emptyset).
\end{align}
When referring to either of these modular functions, we use $m^f_A$.
Note that the $m^f_A$ upper-bound $f$ in the following sense:
$m^f_A(Y) \geq f(Y) \; \forall Y \subseteq V$, and $m^f_A(A) = f(A)$.

\majormin proceeds as follows.  Starting from $A^0 = \emptyset$ and
applying either ``grow'' or ``shrink'' to construct a modular
approximation to $F$ at $\emptyset$ yields the following simple
surrogate function for each $f_i$: $\hat{f_i}(Y) = \sum_{j \in Y}
f_i(j)$.  The below bound then holds
(from~\cite{curvaturesubmodular}):
\begin{equation} \eqlabel{mod-upper-bound}
f_i(Y) \leq \hat{f_i}(Y) \leq \frac{|Y|}{1+(|Y|-1)(1-\kappa_{f_i}(Y))} f_i(Y),\; \forall Y \subseteq V.
\end{equation}

Let $\hat{A} = \argmin_{A \in \mathcal C} \sum_{i =
1}^m \hat{f}_i(A \xor B_i)$. Also, let $A^* = \argmin_{A \in \mathcal
C} \sum_{i = 1}^m f_i(A \xor B_i)$. Then, it holds that:
\begin{align}
\sum_{i = 1}^m f_i(\hat{A} \xor B_i)
  &\leq \sum_{i = 1}^m \hat{f}_i(\hat{A} \xor B_i) \\
  &\leq \sum_{i = 1}^m \hat{f}_i(A^* \xor B_i) \\
  &\leq \frac{|A^*|}{1+(|A^*|-1)(1-\kappa_{f_i}(A^*))} \sum_{i = 1}^m f_i(A^* \xor B_i)
\end{align}
The first inequality follows from the definition of the modular upper
bound, the second inequality follows from the fact that $\hat{A}$ is
the minimizer of the modular optimization, and the third inequality
follows from \eqref{mod-upper-bound}.  We now show that \majormin
improves the objective value at every iteration:
\begin{equation}
\sum_{i = 1}^m f_i(A^{t+1} \xor B_i)
  \leq \sum_{i = 1}^m m^{f_i}_{A^t \xor B_i}(A^{t+1} \xor B_i)
  \leq \sum_{i = 1}^m m^{f_i}_{A^t \xor B_i}(A^t \xor B_i)
  = \sum_{i = 1}^m f_i(A^t \xor B_i).
\end{equation}
\end{proof}}

While \majormin does not have a constant-factor guarantee (which is
possible only in the unconstrained setting), the bounds are not too
far from the hardness of the constrained setting.  For example, in the
cardinality case, the guarantee of \majormin is $\frac{n}{1 + (n-1)(1
- \kappa_f)}$, while the hardness shown in \thmref{cardconst-hard} is
$\Omega\left(\frac{\sqrt{n}}{1 + (n-1)(1 - \kappa_f)}\right)$.

\section{Maximization of the submodular Hamming metric}
\seclabel{maximization}
We next characterize the hardness of \shmax (the diversification
problem) and describe approximation algorithms for it.  We first show
that all versions of \shmax, even the unconstrained homogeneous one,
are NP-hard.  Note that this is a non-trivial result.  Maximization of
a monotone function such as a polymatroid is not NP-hard; the
maximizer is always the full set $V$.  But, for \shmax, despite the
fact that the $f_i$ are monotone with respect to their argument $A
\xor B_i$, they are not monotone with respect to $A$ itself.  This
makes \shmax significantly harder.  After establishing that \shmax is
NP-hard, we show that no poly-time algorithm can obtain an
approximation factor better $3/4$ in the unconstrained setting, and a
factor of $(1 - 1/e)$ in the constrained setting.  Finally, we provide
a simple approximation algorithm which achieves a factor of $1/4$ for
all settings.

\begin{theorem}
All versions of \shmax (constrained or unconstrained, heterogeneous or
homogeneous) are NP-hard.  Moreover, no poly-time algorithm can obtain
a factor better than $3/4$ for the unconstrained versions, or better
than $1 - 1/e$ for the cardinality-constrained versions.
\end{theorem}
\arxiv{\begin{proof}
We first show that homogeneous unconstrained \shmax is NP-hard.  We
proceed by constructing an $F$ that can represent any symmetric
positive normalized (non-monotone) submodular function.  Maximization
is NP-hard for this type of function, since it subsumes
the \textsc{Max-Cut} problem.  Hence, the reduction to
unconstrained \shmax suffices to show NP-hardness.

Consider an instance of \shmax with $m = 2$ and $B_1 = \emptyset, B_2
= V$:
\begin{equation}
\max_{A \subseteq V} F(A) = \max_{A \subseteq V} f(A) +
f(V \setminus A).
\end{equation}
Given a symmetric positive normalized submodular function $h$, define:
\begin{equation}
f(A) = h(A) - \sum_{i \in A} h(i \mid V \setminus i),
\end{equation}
where $h(i \mid V \setminus i)$ is short for $h(V) - h(v \setminus
i)$.  To see that $f$ is a positive polymatroid function, first recall
that a symmetric set function is one for which $h(A) = h(V \setminus
A)\; \forall A \subseteq V$.  Thus, $h(i \mid V \setminus i) = h(V) -
h(V \setminus i) = h(\emptyset) - h(i) = -h(i)$.  This implies that
$f(A) = h(A) + \sum_{i \in A} h(i)$, which is clearly a positive
polymatroid.  Now, notice that:
\begin{equation}
F(A) = f(A) + f(V \setminus A)
  = h(A) + h(V \setminus A) + \sum_{i \in V} h(i)
  = 2h(A) + \sum_{i \in V} h(i).
\end{equation}
Hence, given an instance of symmetric submodular function
maximization, we can transform it into an instance of \shmax, with $F$
defined as above; since $\sum_{i \in V} h(i)$ is a constant, it does
not affect which set is the maximizer.  Thus, unconstrained \shmax is
NP-hard.

To show the hardness of approximation, we borrow a proof technique
from Theorem 4.5 of \cite{MaxNonMonoSubFns}.  The idea is to construct
two symmetric submodular functions, $h_1$ and $h_2$, which
are \emph{indistinguishable}.  That is, any randomized algorithm would
require an exponential number of calls to the value oracles to tell
$h_1$ and $h_2$ apart.  The construction of \cite{MaxNonMonoSubFns}
suggests that for both these functions, $h_1(i) = h_2(i) = n-1$, and
hence the constant $\sum_{i \in V} h(i) = n(n-1)$.  Thus, we can write
$F_1(A) = 2h_1(A) + n(n-1)$ and $F_2(A) = 2h_2(A) + n(n-1)$.  Since
$h_1$ and $h_2$ are indistinguishable, so are $F_1$ and
$F_2$. Moreover, according to \cite{MaxNonMonoSubFns}, the maximum
value of $h_1(A)$ is $n^2/4$, while that of $h_2(A)$ is $n^2/2$.
Hence, $F_1$'s maximum value is $n^2/2 + n^2 - n = 3n^2/2 -n$, and
$F_2$'s is $n^2 + n^2 - n = 2n^2 -n$.  The ratio of these is:
$(3n/2-1)/(2n-1) = 3/4 + o(1)$.  Thus, no poly-time algorithm can
achieve a factor better than $3/4$.

Finally, we establish the hardness of approximation for a
cardinality-constrained version of \shmax.  In this case, let $m = 1$
and $B_1 = \emptyset$.  This \shmax instance is exactly the problem of
monotone submodular maximization subject to cardinality constraint,
which is not only NP-hard but has a hardness of $1
-1/e$~\cite{feige1998threshold}.
\end{proof}}

We turn now to approximation algorithms.  For the unconstrained
setting, \lemref{rand-approx} shows that simply choosing a random
subset, $A \subseteq V$ provides a $1/8$-approximation in expectation.

\begin{lemma} \lemlabel{rand-approx}
A random subset is a $1/8$-approximation for \shmax in the
unconstrained (homogeneous or heterogeneous) setting.
\end{lemma}
\arxiv{\begin{proof}
This result follows from the fact that a random subset is a
$1/4$-approximation for the problem of unconstrained non-monotone
submodular maximization \cite[Theorem 2.1]{MaxNonMonoSubFns}, and that
the non-monotone submodular function $\bar{F}(A)
= \sum_{i=1}^m \left[f_i(A \setminus B_i) + f_i(B_i \setminus
A) \right]$ is within a factor $2$ of the $F(A)$ of \shmax
(see \lemref{Fsplit}).  Thus, a random set is a $1/4$-approximation
for $\max_A \bar{F}(A)$ and a $1/8$-approximation for $\max_A F(A)$.
\end{proof}}

An improved approximation guarantee of $1/4$ can be shown for a
variant of \unionsplit (\algref{union-split}), if the call
to \submodopt is a call to a \submodmax
algorithm.  \thmref{union-split-max} makes this precise for both the
unconstrained case and a cardinality-constrained case.  It might also
be of interest to consider more complex constraints, such as matroid
independence and base constraints, but we leave the investigation of
such settings to future work.

\begin{theorem} \thmlabel{union-split-max}
Maximizing $\bar{F}(A) = \sum_{i=1}^m \left( f_i(A \setminus B_i) +
f_i(B_i \setminus A) \right)$ with a bi-directional greedy
algorithm~\cite[Algorithm 2]{buchbinder2012tight} is a linear-time
$1/4$-approximation for maximizing $F(A) = \sum_{i=1}^m f_i(A \xor
B_i)$, in the unconstrained setting.  Under the cardinality constraint
$|A| \leq k$, using the randomized greedy algorithm~\cite[Algorithm
1]{buchbinder2014submodular} provides a $\frac{1}{2e}$-approximation.
\end{theorem}
\arxiv{
\begin{proof}
$\bar{F}(A)$ is a (non-monotone) submodular function that is within a
factor $2$ of $F(A)$ (see \lemref{Fsplit}).  The bi-directional greedy
algorithm \cite[Algorithm 2]{buchbinder2012tight} provides a
$1/2$-approximation to non-monotone submodular maximization in the
unconstrained setting.  Thus, applying it to $\bar{F}$ yields a
$1/4$-approximation for $\max_A F(A)$.  Similarly, in the
cardinality-constrained setting, one can use the randomized greedy
algorithm~\cite[Algorithm 1]{buchbinder2014submodular}, which has a
$1/e$ approximation guarantee.
\end{proof}}

\section{Experiments}
\seclabel{experiments}
\begin{figure*}
\centering
\begin{minipage}[t]{0.55\textwidth}
\centering
\captionof{table}{mV-ROUGE averaged over the 14 datasets ($\pm$ standard deviation).}
\tablabel{mvrouge}
\begin{tabular}{|c|c|c|}
\hline
\hamming & \onepart & \twopart \\
\hline
$0.38 \pm 0.14$ & $0.43 \pm 0.20$ & $\bf 0.50 \pm 0.26$\\
\hline
\end{tabular}
\end{minipage}
\hspace{0.1in}
\begin{minipage}[t]{0.4\textwidth}
\centering
\captionof{table}{\# of wins (out of 14 datasets).}
\tablabel{mvrougewins}
\begin{tabular}{|c|c|c|} \hline
\hamming & \onepart & \twopart \\ \hline
$3$ & $1$ & $\bf 10$ \\ \hline
\end{tabular}
\end{minipage}
\end{figure*}

\vspace{-1\baselineskip}
To demonstrate the effectiveness of the submodular Hamming metrics
proposed here, we apply them to a metric
minimization task (clustering) and a metric maximization task (diverse
$k$-best).
\vspace{-1\baselineskip}
\subsection{\shmin application: clustering}

We explore the document clustering problem described
in \secref{applications}, where the groundset $V$ is all unigram
features and $B_i$ contains the unigrams of document $i$. We run
$k$-means clustering and each iteration find the mean
for cluster $C_j$ by solving:
\arxivalt{\begin{align} \eqlabel{means-constrained}
\mu_j \in \argmin_{A : |A| \geq \ell} \sum_{i \in C_j} f(A \xor B_i).
\end{align}}{$\mu_j \in \argmin_{A : |A| \geq \ell} \sum_{i \in C_j} f(A \xor B_i).$}
The constraint $|A| \geq \ell$ requires the mean to contain at least
$\ell$ unigrams, which helps $k$-means to create richer and more
meaningful cluster centers.  We compare using the submodular function
$f(Y) = \sum_{W \in \mathcal{W}} \sqrt{|Y \cap W|}$ (\submod), to
using Hamming distance (\hamming).  The problem of finding $\mu_j$
above can be solved exactly for \hamming, since it is a modular
function.  In the \submod case, we apply \majormin
(\algref{major-min}). As an initial test, we generate synthetic data
consisting of $100$ ``documents'' assigned to $10$ ``true'' clusters.
We set the number of ``word'' features to $n = 1000$, and partition
the features into $100$ word classes (the $\mathcal{W}$ in the
submodular function).  Ten word classes are associated with each true
document cluster, and each document contains one word from each of
these word classes.  That is, each word is contained in only one
document, but documents in the same true cluster have words from the
same word classes.  We set the minimum cluster center size to $\ell =
100$.
We use $k$-means++ initialization \cite{arthur2007soda} and average
over $10$ trials.  Within the $k$-means optimization, we enforce that
all clusters are of equal size by assigning a document to the closest
center whose current size is $< 10$.  With this setup, the average
accuracy of \hamming is $28.4$\% ($\pm 2.4$), while \submod is
$69.4$\% ($\pm 10.5$).  The \hamming accuracy is essentially the
accuracy of a random assignment of documents to clusters; this makes
sense, as no documents share words, rendering the Hamming distance
useless.  In real-world data there would likely be some word overlap
though; to better model this, we let each document contain a random
sampling of $10$ words from the word clusters associated with its
document cluster.  In this case, the average accuracy of \hamming is
$57.0$\% ($\pm 6.8$), while \submod is $88.5$\% ($\pm 8.4$).  The
results for \submod are even better if randomization is removed from
the initialization (we simply choose the next center to be one with
greatest distance from the current centers).  In this case, the
average accuracy of \hamming is $56.7$\% ($\pm 7.1$), while \submod is
$100$\% ($\pm 0.0$).  This indicates that as long as the starting
point for \submod contains one document from each cluster, the \submod
optimization will recover the true clusters.\looseness-1

Moving beyond synthetic data, we applied the same method to the
problem of clustering NIPS papers.  The initial set of documents that
we consider consists of all NIPS papers\footnote{Papers were
  downloaded from \url{http://papers.nips.cc/}.} from 1987 to 2014.
We filter the words of a given paper by first removing stopwords and
any words that don't appear at least $3$ times in the paper.  We
further filter by removing words that have small tf-idf value ($<
0.001$) and words that occur in only one paper or in more than 10\% of
papers.  We then filter the papers themselves, discarding any that
have fewer than $25$ remaining words and for each other paper
retaining only its top (by tf-idf score) $25$ words.  Each of the
$5{,}522$ remaining papers defines a $B_i$ set.  Among the $B_i$ there
are $12{,}262$ unique words.  To get the word clusters $\mathcal{W}$,
we first run the \textsc{word2vec} code of \cite{mikolov2013nips},
which generates a $100$-dimensional real-valued vector of features for
each word, and then run $k$-means clustering with Euclidean distance
on these vectors to define $100$ word clusters.  We set the center
size cardinality constraint to $\ell = 100$ and set the number of
document clusters to $k = 10$.  To initialize, we again use
$k$-means++ \cite{arthur2007soda}, with $k = 10$.  Results are
averaged over $10$ trials. While we do not have groundtruth labels for
NIPS paper clusters, we can use within-cluster distances as a proxy
for cluster goodness (lower values, indicating tighter clusters, are
better).  Specifically, we compute: $k\textrm{-means-score} = \sum_{j
  = 1}^k \sum_{i \in C_j} g(\mu_j \xor B_i)$.  With Hamming for $g$,
the average ratio of \hamming's $k$-means-score to \submod's is $0.916
\pm 0.003$.  This indicates that, as expected, \hamming does a better
job of optimizing the Hamming loss.  However, with the submodular
function for $g$, the average ratio of \hamming's $k$-means-score to
\submod's is $1.635 \pm 0.038$.  Thus, \submod does a significantly
better job optimizing the submodular loss.\looseness-1

\vspace{-1\baselineskip}
\subsection{\shmax application: diverse $k$-best}
\vspace{-1\baselineskip}

In this section, we explore a diverse $k$-best image collection
summarization problem, as described in~\secref{applications}.  For this
problem, our goal is to obtain $k$ summaries, each of size $l$, by
selecting from a set consisting of $n \gg l$ images.  The idea is that
either: (a) the user could choose from among these $k$ summaries the
one that they find most appealing, or (b) a (more computationally
expensive) model could be applied to re-rank these $k$ summaries and
choose the best.  As is described in \secref{applications}, we obtain
the $k$th summary $A_k$, given the first $k-1$ summaries $A_{1:k-1}$
via: $A_k = \argmax_{A \subseteq V, |A| =
  \ell} g(A) + \sum_{i = 1}^{k-1} f(A \xor A_i)$.
\begin{wrapfigure}{r}{0.5\textwidth}
\vspace{-2ex}
\begin{center}
\includegraphics[width=0.5\textwidth]{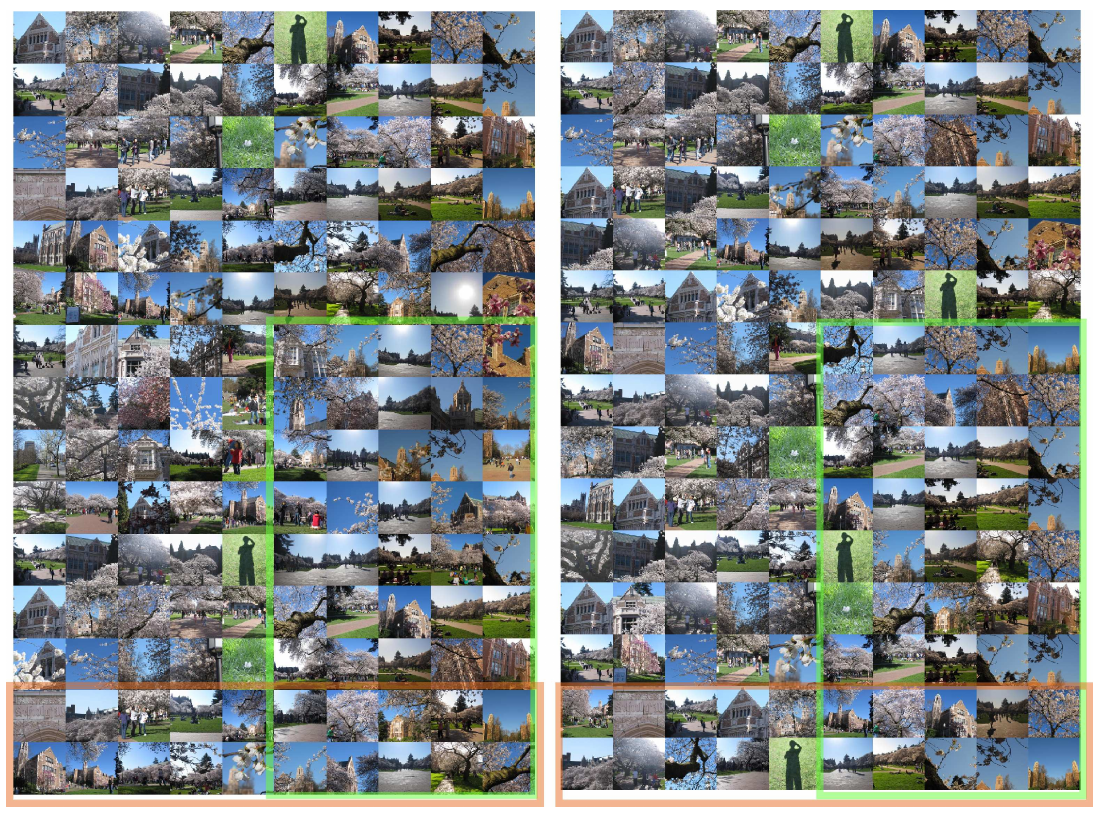}
\end{center}
\vspace{-3ex}
\caption{An example photo montage (zoom in to see detail) showing $15$ summaries of size $10$ (one per row) from the \hamming approach (left) and the \twopart approach (right), for image collection \#6.}
  \figlabel{imgsumm}
\vspace{-2ex}
\end{wrapfigure}
For $g$ we use the facility location function: $g(A) = \sum_{i \in V}
\max_{j \in A} S_{ij}$, where $S_{ij}$ is a similarity score for
images $i$ and $j$.  We compute $S_{ij}$ by taking the dot product of
the $i$th and $j$th feature vectors, which are the same as those used
by~\cite{tschiatschek2014learning}.  For $f$ we compare two different
functions: (1) $f(A \xor A_i) = |A \xor A_i|$, the Hamming distance
(\hamming), and (2) $f(A \xor A_i) = g(A \xor A_i)$, the submodular
facility location distance (\submod).  For \hamming we optimize via
the standard greedy algorithm~\cite{nemhauser1978mp}; since the
facility location function $g$ is monotone submodular, this implies an
approximation guarantee of $(1 - 1/e)$.  For \submod, we experiment
with two algorithms: (1) standard greedy \cite{nemhauser1978mp}, and
(2) \unionsplit (\algref{union-split}) with standard greedy as the
\submodopt function.  We will refer to these two cases as ``single
part'' (\onepart) and ``two part'' (\twopart).  Note that neither of
these optimization techniques has a formal approximation guarantee,
though the latter would if instead of standard greedy we used the
bi-directional greedy algorithm of \cite{buchbinder2012tight}.  We opt
to use standard greedy though, as it typically performs much better in
practice.  \jenny{Jeff}{Ideally we'd use both and take the best of
  greedy and bi-directional greedy.}  We employ the image
summarization dataset from~\cite{tschiatschek2014learning}, which
consists of $14$ image collections, each of which contains $n = 100$
images. For each image collection, we seek $k = 15$ summaries of size
$\ell = 10$.  For evaluation, we employ the V-ROUGE score developed
by~\cite{tschiatschek2014learning}; the mean V-ROUGE (mV-ROUGE) of the
$k$ summaries provides a quantitative measure of their goodness.
V-ROUGE scores are normalized such that a score of $0$ corresponds to
randomly generated summaries, while a score of $1$ is on par with
human-generated summaries.

\tabref{mvrouge} shows that \onepart and \twopart outperform \hamming
in terms of mean mV-ROUGE, providing support for the idea of using
submodular Hamming distances in place of (modular) Hamming for diverse
$k$-best applications.  \twopart also outperforms \onepart,
suggesting that the objective-splitting used in \unionsplit is of
practical significance.  \tabref{mvrougewins} provides additional
evidence of \twopart's superiority, indicating that for $10$ out of
the $14$ image collections, \twopart has the best mV-ROUGE score of
the three approaches.\looseness-1

\figref{imgsumm} provides some qualitative evidence of \twopart's
goodness.  Notice that the images in the green rectangle tend to be
more redundant with images from the previous summaries in the \hamming
case than in the \twopart case; the \hamming solution contains many
images with a ``sky'' theme, while \twopart contains more images with
other themes.  This shows that the \hamming solution lacks diversity
across summaries.  The quality of the individual summaries also tends
to become poorer for the later \hamming sets; considering the images
in the red rectangles overlaid on the montage, the \hamming sets
contain many images of tree branches here.  By contrast, the \twopart
summary quality remains good even for the last few
summaries.\looseness-1

\vspace{-0.5\baselineskip}
\section{Conclusion}
\seclabel{conclusion}
\vspace{-0.5\baselineskip}
In this work we defined a new class of distance functions: submodular
Hamming metrics.  We established hardness results for the associated
\shmin and \shmax problems, and provided approximation algorithms.
Further, we demonstrated the practicality of these metrics for several
applications.  There remain several open theoretical questions (e.g.,
the tightness of the hardness results and the NP-hardness of \shmin),
as well as many opportunities for applying submodular Hamming metrics
to other machine learning problems (e.g., the \arxiv{structured}
prediction application from \secref{applications}).



\subsubsection*{References}

{\small
\bibliographystyle{unsrt}
\begingroup
\renewcommand{\section}[2]{}
\bibliography{paper}
\endgroup
}

\end{document}